# Performance-Enhanced Non-Enzymatic Glucose Sensor Based on Graphene-Heterostructure

**Mahmoud A. Sakr [1,2], Karim Elgammal [3,4], Anna Delin [3,4,5] and Mohamed Serry [2,\*]**

1. Graduate Program in Nanotechnology, The American University in Cairo (AUC), New Cairo 11835, Egypt; mahmoud.a.sakr@aucegypt.edu
2. Department of Mechanical Engineering, The American University in Cairo (AUC), New Cairo 11835, Egypt
3. Department of Applied Physics, School of Engineering Sciences, KTH Royal Institute of Technology, Electrum 229, SE-16440 Kista, Sweden; elgammal@kth.se (K.E.); annadel@kth.se (A.D.)
4. Swedish e-Science Research Center (SeRC), KTH Royal Institute of Technology, SE-10044 Stockholm, Sweden
5. Department of Physics and Astronomy, Materials Theory Division, Uppsala University, Box 516, SE-75120 Uppsala, Sweden
* Correspondence: mserry@aucegypt.edu; Tel.: +202-2615-3098

**Abstract:** Non-enzymatic glucose sensing is a crucial field of study because of the current market demand. This study proposes a novel design of glucose sensor with enhanced selectivity and sensitivity by using graphene Schottky diodes, which is composed of graphene (G)/platinum oxide (PtO)/n-silicon (Si) heterostructure. The sensor was tested with different glucose concentrations and interfering solutions to investigate its sensitivity and selectivity. Different structures of the device were studied by adjusting the platinum oxide film thickness to investigate its catalytic activity. It was found that the film thickness plays a significant role in the efficiency of glucose oxidation and hence in overall device sensitivity. 0.8–2 µA output current was obtained in the case of 4–10 mM with a sensitivity of 0.2 µA/mM.cm². Besides, results have shown that 0.8 µA and 15 µA were obtained by testing 4 mM glucose on two different PtO thicknesses, 30 nm and 50 nm, respectively. The sensitivity of the device was enhanced by 150% (i.e., up to 30 µA/mM.cm²) by increasing the PtO layer thickness. This was attributed to both the increase of the number of active sites for glucose oxidation as well as the increase in the graphene layer thickness, which leads to enhanced charge carriers concentration and mobility. Moreover, theoretical investigations were conducted using the density function theory (DFT) to understand the detection method and the origins of selectivity better. The working principle of the sensors puts it in a competitive position with other non-enzymatic glucose sensors. DFT calculations provided a qualitative explanation of the charge distribution across the graphene sheet within a system of a platinum substrate with D-glucose molecules above. The proposed G/PtO/n-Si heterostructure has proven to satisfy these factors, which opens the door for further developments of more reliable non-enzymatic glucometers for continuous glucose monitoring systems.

## 1. Introduction

Electrochemical biosensors work through the perception of a chemical reaction that takes place between an electrode and the bio-analyte of interest to transduce an electrical signal. Thus, they consist of two main elements; a biological sensing probe (i.e., recognition elements) and a transducer. For the latter, amperometry, voltammetry, and electrochemical impedance spectroscopy (EIS) are the most common transduction methods in electrochemical biosensors, in which detection is based on either electron-transfer rate, background current, or accumulation of analytes.

Since its inception in 1964 [1], electrochemical biosensing has been the most widely researched and commercialized method for blood glucose and bio-analyte monitoring. Their primary advantages over other biosensing methods (e.g., optical, thermal, and FET biosensors) are their ease of fabrication and their ability to detect bio-analytes without interfering with the biological system. The advance of electrochemical biosensing has leaped since the introduction of nanomaterials and bioMEMS in the late 1990s which stimulated enhancing electrochemical biosensing performance (e.g., detection limit, sensitivity, and signal-to-noise ratio) through more advanced device assemblies and the integration of nanomaterials-based bio-interface materials (e.g., metal nanoparticles, and carbon nanotubes) as electrode surface modifiers [1–4].

Electrochemical glucose monitoring biosensors (i.e., glucometers) are of particular interest to this study. They have been the most growing medical devices with market size of \$12.8 billion in 2019 [5]. Current research on electrochemical glucometers focuses on enhancing their reliability, stability, sensitivity, and portability while maintaining low cost [6]. In principle, there are two electrochemical glucose detection approaches; enzymatic and non-enzymatic. Enzymatic approaches employ glucose oxidase enzyme (GOx) as a recognition element. They work by oxidizing glucose molecules and the generation of detectable compounds such as $O_2$, $CO_2$, or $H_2O_2$. In this aspect, glucose oxidase breaks into gluconolactone, hydrogen, and oxygen when interacting with enzymatic sensors. The detection mechanism of the sensors depends on the concentrations of the original glucose on the surface of the sensor, which results in the oxidizing of glucose and presenting in typical current values in response to the concentration of the glucose. Although they have high specificity, which ultimately reflects in the overall device selectivity, there are still significant challenges in the development of GOx-based glucometers, namely (i) their sensitivity is degraded with time due to enzyme leaching; (ii) they suffer from low stability and short life-time; being a protein, GOx enzyme highly susceptible to environmental factors such as pH, chemical reagents, temperature, and humidity; and (iii) reduction in high overpotentials.

Hence, these challenges have stimulated the development of more reliable electrochemical glucometers through an alternative non-enzymatic approach. In this approach, a more stable and reliable metal or metal oxide-based electrocatalytic media (e.g., Pt or CuO) is employed to oxidize glucose on the electrode surface directly. Several developments of non-enzymatic have reported enhanced sensitivity (e.g., 0.01–0.46 $\mu A\ mM^{-1}\ cm^{-2}$ [7,8], fast response (e.g., 1 s, [9]), and low-cost and high reproducibility ([10,11])). However, when compared to enzymatic sensors, most of the reported sensors suffered from a short linear range (e.g., 0.5–7.5 mM, [12]), and low selectivity [12] due to the interference with other chemical components in whole mammalian blood such as ascorbic or uric acids. Hence, the most sought-after achievement for the next generation non-enzymatic glucometers is the improvement in the device selectivity, reliability, and sensitivity [13].

As such, several technologies have been investigated to develop the electrochemical reaction-based electrode surfaces in non-enzymatic glucometers including carbon-based materials (e.g., graphene, rGO, and CNT) [14,15], and metal nanoparticles (e.g., gold) [16,17], which due to their remarkable electrical and physical/chemical properties, they have been widely investigated as electrochemical electrodes in various electrochemical biosensor applications [18,19].

Accordingly, in this work, we have investigated and optimized the physio/chemical performance of a novel G/PtO/n-Si heterostructured electrode for non-enzymatic glucose biosensors with the primary objective of enhancing the linear detection range and reducing the interference between glucose and other chemicals in the whole blood in order to enhance selectivity. The critical challenge in this research is to achieve high selectivity while maintaining, wide linear range and low detection limit; this is carried out by optimizing the G/PtO film thickness and enhancing their interface. Therefore, for the first time, an ALD-grown platinum oxide (PtO) was effectively employed as both a catalyst to grow graphene and an electrocatalytic agent for the oxidation of glucose and $H_2O_2$ molecules, respectively. Pt and PtO are considered as the strongest catalytic metals/metal oxides for glucose detection [20–22]. However, there are some limitations associated with Pt/PtO sensors, including their degradation due to the adsorption of physiological solutions such as some uric and amino acids [11,12,17,23]. Bai et al. [18] hypothesized the advantage of using $Pt_xCO_{1-x}$ alloy nanoparticles and carbon as a support, with this alloy representing the enhancement of device selectivity toward glucose and linear response at different concentrations. Moreover, it avoids the competitive effects of fructose, uric acid, and ascorbic acid. Hence, in this work, graphene is also employed as both a protective and sieving layer to protect the PtO film and further enhance sensor stability and selectivity. In addition, due to its high electrical conductivity, graphene-based electrodes contribute to the enhancement of device sensitivity.

Theoretical and experimental analyses revealed that G/PtO interface results in a carrier-free region where the two materials align in their fermi levels by charge transfer from graphene to Pt [21]. Graphene is p-doped because of their Fermi-level alignment, and from I–V characteristics, it demonstrates when applying a forward or reverse bias, it forms symmetric current because of electrons flow [24–26]. Besides, a depletion region is formed at the interface of PtO/Si due to the migration of carriers, which depends on the carrier distribution function and the concentration at the silicon surface.

We have further investigated the selectivity and charge transfer behavior using a density functional theory (DFT) approach. The model findings were in line with the experimental results as the charge transfer of D-type glucose adsorbed on top of a graphene sheet resting on top of a Pt <111> substrate cut was observed, as well as the direct adsorption of D-glucose on top of a Pt <111> substrate surface.

Sensitivity enhancement by ~70–150% was observed by altering PtO and G layer thicknesses. All sensors structures have exhibited excellent selectivity when examined against whole blood components such as sodium chloride (NaCl), potassium chloride (KCl), urea extra pure, potassium, and phosphate monobasic ($K_2PO_4$). From this work, the sensor showed an enhanced sensitivity in the by changing the thickness of the PtO layer, as shown in the hysteresis analysis, selectivity, and cyclic voltammetry figures.

## 2. Materials and Methods

### 2.1. Chemicals

D-(+)-glucose monohydrate ($C_6H_{12}O_6 \cdot H_2O$), sodium phosphate monobasic dehydrate ($H_6NaO_6P$), sodium chloride (NaCl), and potassium chloride (KCl) were purchased from Sigma-Aldrich. Urea extra pure was obtained from Labchemie. Potassium phosphate monobasic $K_2PO_4$ was obtained from Aldrich Chem. Co. Phosphate buffer saline (pH = 7.4) 0.1 mM was prepared fresh for every experiment.

### 2.2. Device Fabrication

Starting from an N-type silicon wafer and then the deposition of a PtO thin film was performed by atomic layer deposition (ALD) using Cambridge NanoTech Savannah ALD deposition system. The used precursors were Trimethyl (methylcyclopentadienyl) Pt (IV) with high purity $O_2$ used as an oxidizing agent. Deposition temperature was 275 °C for 50–500 ALD-cycles to produce a PtO thin film with thickness ranging from 50 to 500 Å, respectively. Subsequently, graphene bottom-up growth was done in Plasma Enhanced Chemical Vapor Deposition (PECVD) using Oxford Instruments, Plasmalab 100 PECVD System. The pressure was maintained at 1500 mTorr, and the chamber was heated from 195 to 600 °C in a methane ($CH_4$) rich environment. High frequency 13.56 MHz and low frequency 380 kHz powers were varied to control the structure and thickness of the obtained graphene structure.

### 2.3. Structural Characterization

Morphological characterization was obtained with a Field Emission Scanning Electron Microscope (FESEM) by Zeiss. X-ray photo-electron spectroscopy (XPS) characterization data were performed on K-alpha XPS from Thermo Scientific in the range of 1–1350 eV to inspect the surface chemistry of the graphene. Al anode was used, whereas the spectra were charge referenced to O1s at 532 eV. The crystalline phases were detected and identified by a glancing angle X-ray diffractometer (GAXRD) on an X'pert PRO MRD with copper source. Raman spectra of the graphene thin film were obtained by Enwave Optronics Raman Spectroscopy ($\lambda_{ex}$ = 532 nm, P = 500 mW, acquisition time = 10 s).

*2.4. Testing Setup and Theoretical Modeling*

The device was tested via a semiconductor parameter analyzer for current–voltage (I–V) and current–time (I–t) measurements to verify the working mechanism by using different glucose concentrations and observing the effect of different competitive solutions as discussed in Section 3.4.

Different glucose concentrations (0.1–1 mM) in phosphate buffer saline (PBS) (pH = 7.4) were tested on a biologic potentiostat for cyclic voltammetry, which was conducted at (−1.5:1.5) mV with scan rate 20 mV. Measurements included three-electrode cells with Ag/AgCl as a reference electrode and a Pt sheet as a counter electrode. Nitrogen gas purging was carried out during the process to keep the surface active and stop any formation of oxide layers.

The electrochemical analysis was obtained using Biologic Potentiostat for cyclic voltammetry and amperometry measurements to identify oxidation and reduction of glucose molecules on the electrode surface, as well as the electrochemical electron transfer process of glucose on the surface. In an attempt to investigate the change in charge distribution due to the presence of glucose molecules on the surface, DFT using Quantum ESPRESSO was helpful in solving the Schrodinger equation and for studying charge density in the presence and absence of glucose.

## 3. Results and Discussion

*3.1. Working Principle*

The working principle of the device as shown in Figure 1 depends on the oxidation of glucosemolecules on the surface of the proposed heterostructured electrode due to the catalytic activity of PtO thin film. Besides, the breakdown of glucose molecules into gluconolactone and the formation of gluconic acid, hydrogen $H_2$, and electrons [27]. Forward bias was applied to the Si terminal, and the current was measured from the graphene surface. The physisorption interaction between graphene and PtO (i.e., metal-semiconductor Schottky junction) resulted in a shift in the fermi-level position and p-doping of graphene.

PtO was chosen as a suitable candidate for two main reasons: its catalytic activity towards graphene growth and its catalytic activity towards glucose oxidation [25,28,29]. Pt catalytic activity towards graphene growth was studied previously by the authors owing to the intense catalytic activity of PtO for $CH_4$ dissociation compared to Cu or Ni, and for its strong ability for $H_2$ dissociation [26,27,30].

Whereas, the electrocatalytic oxidation of D-glucose at the PtO surface is proposed to take multiple steps as follows: graphene is considered an excellent supporter of electrochemical devices, as it enhances electrical conductivity, selectivity, stability, and sensitivity [2]. Oxidation and reduction of glucose molecules result in the formation of gluconolactone and the release of electrons and $H_2$, and then the formation of gluconic acid [2]. Finally, the electrons from oxidation will start to cause variations in the Schottky barrier height.

As such, oxidation of glucose molecules on the surface structure changed the local electric field distribution and the variation in Schottky barrier height (SBH) in PtO/n-Si, resulting in detectable current changes due to molecule adsorption. The working principle of the structure in the presence and absence of glucose molecules on the device. As shown, in the absence of glucose, the fermi level of the graphene layers was aligned with the fermi level of PtO. With the presence of glucose molecules, fermi levels were aligned. After applying bias on the silicon, the fermi levels started to shift due to the oxidation of glucose due to the liberating free electrons as a result of oxidation. This shift was noticed in the Schottky barrier height between G/PtO and Si. The shift in Fermi levels and the change in the barrier height can be correlated to the concentration of glucose with significant linearity.

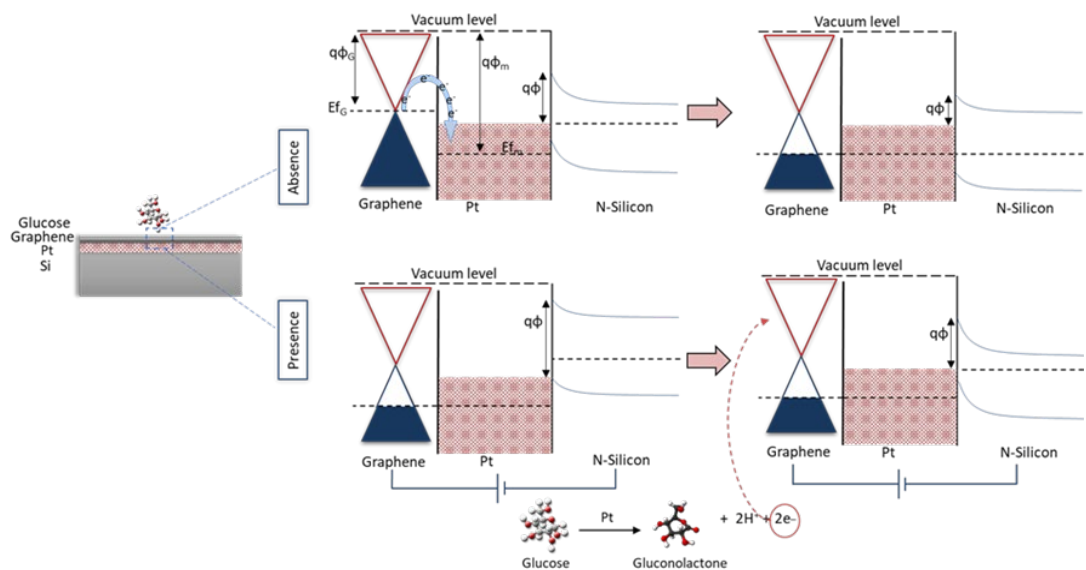

**Figure 1.** Graphical scheme for the working principle of the device.

*3.2. Theoretical Study*

To gain more insight into the adsorption of D-glucose on top of graphene residing on a slab of PtO, density functional theory (DFT) simulations were performed. The PtO slab was formed by a <111> surface cut in a bulk PtO cubic system, forming a hexagonal substrate. The DFT calculations were carried out using the plane-wave basis set Quantum ESPRESSO (QE) [31] code. The optimized norm-conserving scalar relativistic [29,32] pseudopotentials (ONCV) were downloaded from PseudoDojo and developed by the University of California were used at the recommended plane-wave cut-off of 60 Ry. The generalized gradient approximation (GGA) functional was used with the Perdew, Burke, and Ernzerhof (PBE) [33] parameterization for the exchange-correlation part in conjunction with the semi-empirical Grimme correction [34,35] with implementation in QE as a correction for the weak van der Waals (vdW) forces. Gaussian smearing of 0.03 Ry was applied. The Brillouin zone was sampled with 9 × 9 × 1 k-point mesh when the supercell was folded to the graphene's primitive unit cell. The calculational supercell had a vacuum set to ≈ 20 Å to eliminate any interaction between supercell repetitions in the z-direction. The slab cuts, as well as the supercell formation and transformation operations, were generated via the CIF2 Cell utility code [36,37] and the Virtual NanoLab version 2016.1 from Quantum Wise A/S. The D-glucose structure was downloaded as an XYZ formatted structure file [38,39].

*3.3. Structural Characterization*

Figure 2a illustrates the XPS survey spectra of the graphene surface after 0, 50, and 100-sec etching of the surface. Three main peaks were resolved from this figure with binding energies of 75.08 eV for Pt, in the range of 284–285 eV for C, our measurements were obtained in 1 eV increment; therefore the resolution of our measurement did not allow us to capture the correct range from 284.5 to 284.7 eV. Whereas, for O1s, the binding energies were detected at 532 eV, 525 eV, and 522 eV at 0, 50, and 100 s etching respectively [40,41]. It is observed that the peaks obtained indicating the existence of both graphene and PtO near the surface, which confirms the electrocatalytic/charge transfer detection principle outlined in the previous section.

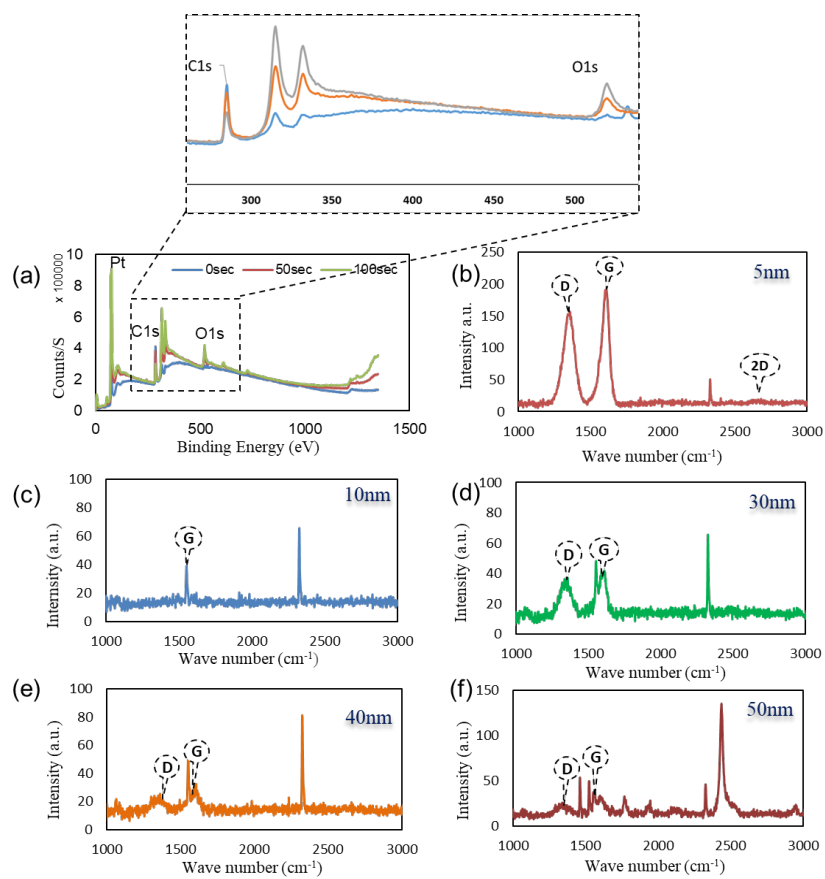

**Figure 2.** (**a**) XPS survey spectra on the surface of the graphene film after 0 s, 50 s, and 100 s etching of the surface. RAMAN spectra of graphene layers on top of different Pt thicknesses (**b**) 5; (**c**) 10; (**d**) 30; (**e**) 40; (**f**) 50 nm film.

Figure 2b–f illustrates the Raman spectrum of the graphene surface deposited at various deposition conditions with PtO film layer thickness of 5, 10, 30, 40, and 50 nm. The figure displays the graphene's D, G, and 2D bands. The shift in the G band can be attributed to the PtO/G physisorption interaction [42]. Bonding between graphene and metallic atoms might affect the work function of graphene and the electronic band structure. In our case, the physisorption interaction between PtO/G might result in a slight shifting in graphene's Fermi level. In our case, the work function of PtO is 5.98 eV and graphene's is 4.48 eV [43,44].This interaction can be noticed in the shift in the Raman peaks[45,46]. Accordingly, shifts in the Raman D, G, and 2D bands in 1324 cm$^{-1}$, 1574 cm$^{-1}$, 2692 cm$^{-1}$ were observed, respectively while the peak at 2400 cm$^{-1}$ is a combination between the D and D" peaks and in literature, it mentioned to be G* mode [47,48]. Raman is useful in counting the number of graphene layers, and this could be done by the line shape of the 2D peak or the intensities ratio of G/2D and G/Si peaks for CVD graphene synthesis [33,49]. Graphene thickness could be a factor of enhancing the performance of the device due to the uniform charge distribution of the free electrons from the oxidation process. As discussed earlier in our previous work, fabrication of graphene using CVD is affected by the catalyst layer in our case Pt and by increasing the thickness of Pt substrate, this enhanced and modulated graphene layers formed on the surface of the catalyst. Charge transfer in case of 50 nm PtO shows higher sensitivity of the device due to (1) the higher catalytic activity of the device and (2) enhancement in charge transfer due to the increasing number of graphene layers [33,34]. Figure 3a–e FESEM was employed to investigate the morphology of PtO thin films on the n-Si surface and to measure its thickness. As shown in Figure 3, the orientation of PtO atoms on the surface of n-Si and the enhancement of the nucleation/growth of the PtO on the surface was directly related to the number of ALD cycles. A thin layer of PtO is formed on the surface of n-Si. In addition, as shown in Figure 3f, the thickness of the PtO thin film is in the order of 50 nm at 500 ALD-cycles.

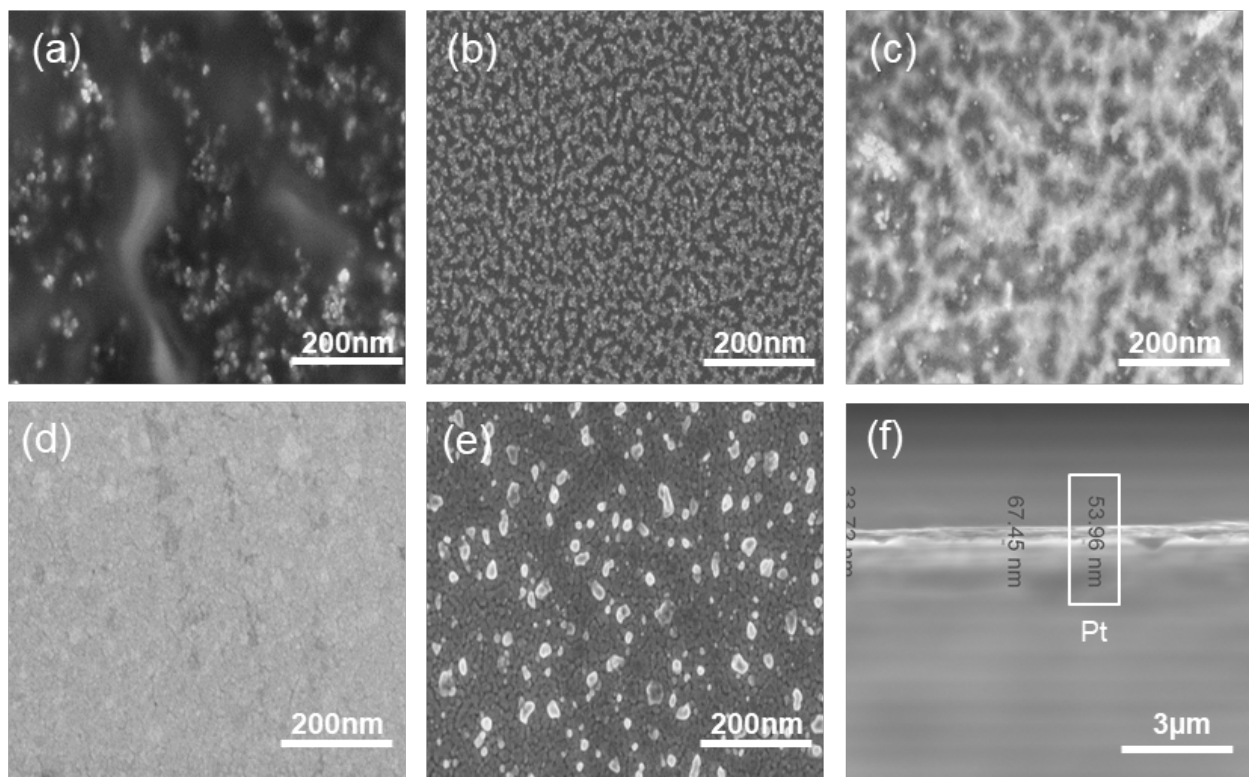

**Figure 3.** FESEM of graphene layers on top of different Pt thicknesses (**a**) 5; (**b**) 10; (**c**) 30; (**d**) 40; (**e**) 50 nm; (**f**) cross section of 50 nm PtO.

*3.4. In Situ Sensitivity, Selectivity, and Linearity Tests*

Using a graphene Schottky diode, we tested different glucose concentrations to obtain the sensitivity of the device. The semiconductor parameter analyzer setup was used to investigate the response of the diode in the circuit to measure the charge difference due to glucose oxidation and the formation of gluconolactone and the subsequent $H_2$ adsorption on the graphene surface, as well as the charge transfer in the opposite directions under forward bias. Moreover, a significant correlation was noticed between the output current and the concentration of the glucose. A 0.8–2 µA difference was noticed in the case of 4–10 mM glucose, as shown in Figure 4a. In addition, the device showed no response in the presence of deionized water or towards the interferent analytes that are typically found in human serum, including sodium chloride, and urea which clarifies the selectivity of the device towards glucose molecules. From these results, the sensitivity of the device is 0.2 µA/mM.cm², which is comparable to recent sensitivity values in literature, as discussed in Section 3.5. While the device sensitivity is moderate as compared to the present carbon-based materials, the selectivity and linearity figures are among the highest in the non-enzymatic class.

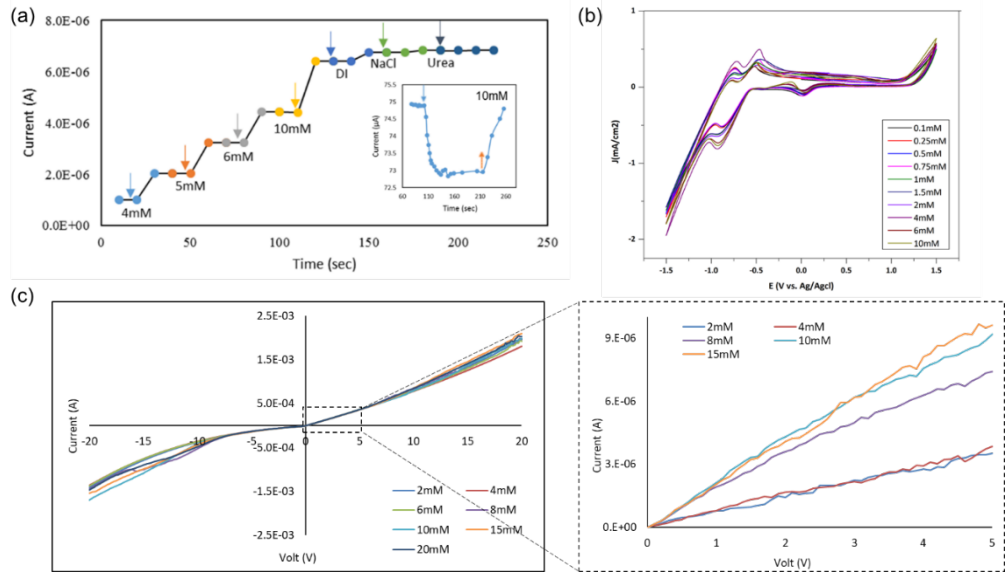

**Figure 4.** (**a**) Testing 4–10 mM of glucose on G/30 nm PtO/n-Si; (**b**) cyclic voltammetry of different glucose concentrations (0.1–10 mM) using a G/30 nm PtO; (**c**) I–V characteristics of the presence of different glucose concentrations from 2 to 20 mM at a sweep voltage from −20 to 20 V.

As shown in Figure 4b from cyclic voltatmetry, the oxidation and reduction peaks confirmed the formation of gluconolactone, gluconic acid molecules, free electrons, and dissociated H2 on the surface of the device. I–V characteristics of the device show a potential difference between the different glucose concentrations from 2 to 20 mM on 30 nm PtO. It shows consistency and significant difference, as shown in Figure 4c. In addition, the I–V characteristics of different glucose concentrations 2–20 mM show the linearity of the response with a sweep bias between 5 and −5 V.

A selectivity test was obtained using a potentiostat and a parameter analyzer to test a 2.4 × 3 cm$^2$ sensor under 0.3 V bias in bare PBS and after adding different chemical solutions (i.e., $H_2O_2$, KCl, NaCl, $Na_2SO_4$, and urea). It was noticed there was no response from the device in the presence of the different solutions, and there was no blocking of the active sites, as shown in Figure 5c. In addition, illustrates the I–V characteristics of the structure in the presence of different solutions to check the selectivity. The results show that the responses of NaCl and urea, separate and together, are very small compared to the glucose readings.

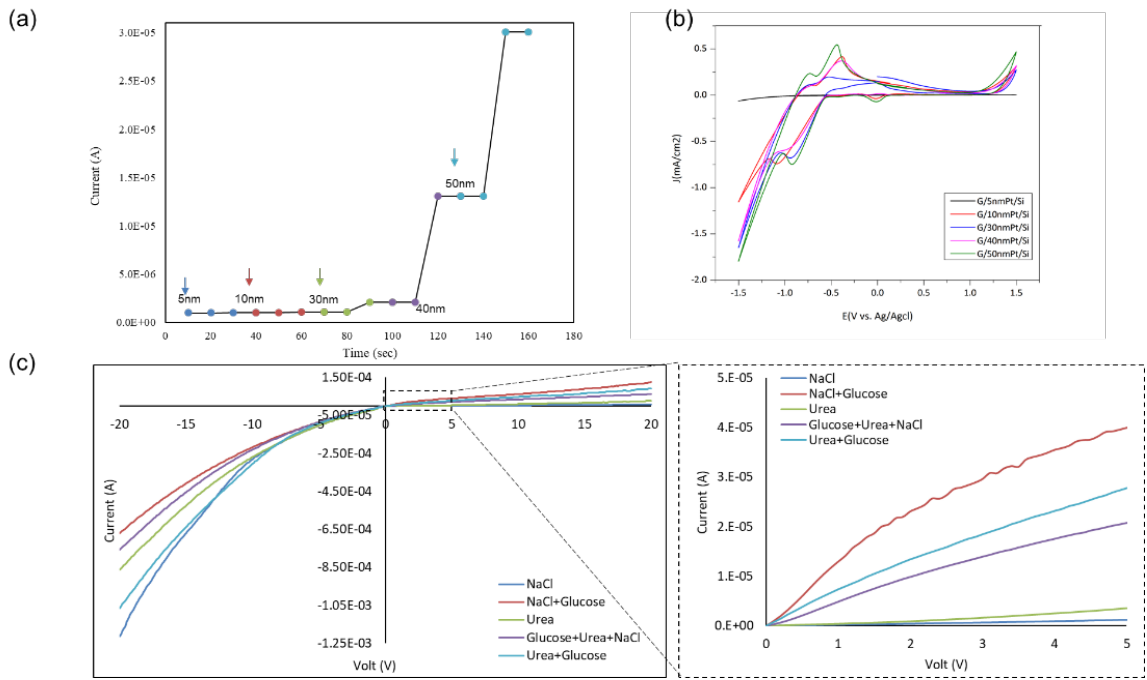

**Figure 5.** Testing 4 mM glucose on (**a**) testing 4 mM on different PtO thicknesses (5–50 nm), (**b**) cyclic voltammetry of different PtO thinkesses at a sweep −1.5–1.5 V; (**c**) testing DI, NaCl, and urea characteristics of the structure in the presence of NaCl, urea, glucose, and mixtures.

*3.5. Effect of PtO Thickness on Device Performance*

Varying the PtO film thickness resulted in an enhancement in graphene growth due to the enhanced nucleation [33,34]. Device sensitivity and its correlation to PtO film thickness were studied by testing different PtO thicknesses with a fixed glucose concentration of 4 mM, it was noticed that there was a strong correlation between the thickness of PtO which the enhanced catalytic activity of the device by increasing the oxidation rate and hence increasing output current/sensitivity, as shown in Figure 5a,b. Enhancement of the sensitivity in the order up to ~150% was observed as the PtO film thickness was increased from 5 to 50 nm, respectively. Accordingly, it is observed that the sensitivity of the device is 0.2 µA/mM.cm$^2$ in case of 4 mM glucose with 30 nm PtO and by increasing the thickness of PtO, the sensitivity will increase up to 30 µA/mM.cm$^2$ which is exceeding the ranges reported in the case of using platinum graphene hybrid electrodes for non-enzymatic glucose detection with a sensitivity of 11 µA/mM.cm$^2$ [30] and platinum nickel alloy nanoparticles with graphene for non-enzymatic detection of glucose with a sensitivity of 20.42 µA/mM.cm$^2$ [50,51].

*3.6. Device Reliability and Hysteresis Analysis*

Figure 6a,b illustrates the average reading of the device as 2 µA for 10 mM and 1 µA for 5 mM glucose, where the blue arrows represent adding glucose on the sensor, and the red ones for removing glucose microdroplets. Figure 6c shows the usability of the sensor for 12 rounds for 1.15 h with a standard deviation of 0.5%, which makes the device suitable for long-term studies.

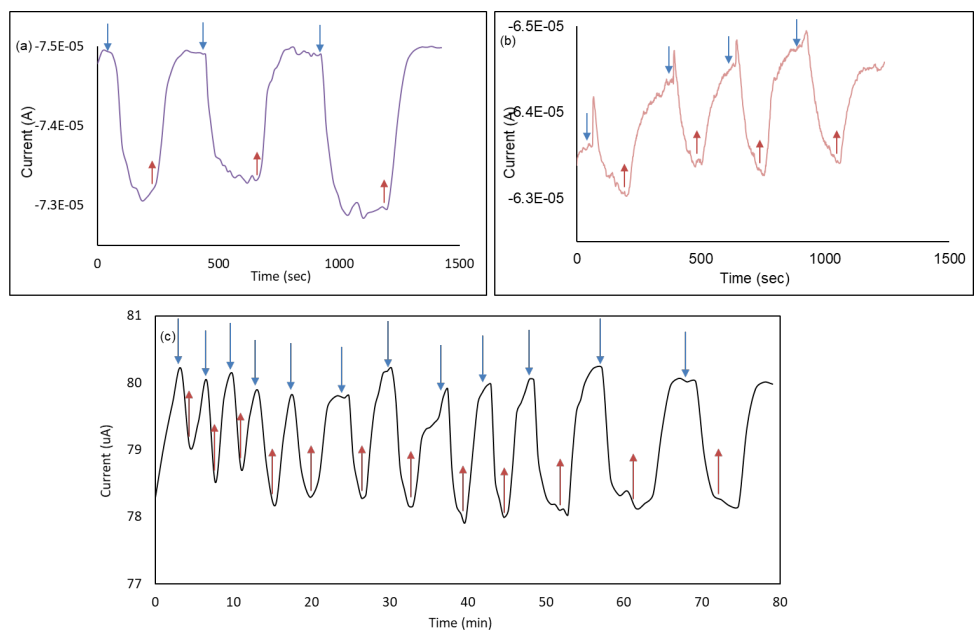

**Figure 6.** Response and hysteresis analysis. (**a**) I–t measurements for 10 mM on G/30 nm PtO/n-Si device; (**b**) I–t measurements for 5 mM on G/30 nm PtO/n-Si device; (**c**) Reproducibility test for 1 h 20 min using 10 mM glucose on the same structure.

*3.7. Theoretical Analysis*

The sensing properties of graphene have been investigated for different species adsorbed on its surface [52–56]. Graphene's electronic structure properties and response to adsorbates were proven to change depending on the underlying substrate, as well as adsorbates existing on top [42–44,57]. The functionalized graphene response to D-glucose adsorbates on top was examined recently towards building a glucose biosensor device. Smith et al. [58–61] investigated graphene's humidity and $CO_2$ sensing behavior through visualizing charge transfer throughout the graphene sheet within the calculational supercell, where they analyzed the effect of water and $CO_2$ adsorbates on top of a graphene sheet. Similarly, in this work, the charge transfer of D-type glucose adsorbed on top of a graphene sheet resting on top of a Pt <111> substrate cut was noticed, as well as the direct adsorption of D-glucose on top of a Pt <111> substrate surface.

Calculations were performed on two calculational setups, where the first was graphene relaxed on top of a Pt <111> substrate surface with the D-glucose structure relaxed on top of the graphene sheet, while the second was the Pt <111> substrate with the D-glucose structure relaxed directly on top of the Pt <111> substrate. All system components were relaxed where the graphene sheet recorded a spacing of 3.1 Å above the Pt substrate, while the D-glucose recorded 3.5 Å above the graphene sheet and 3.63 Å above the Pt <111> substrate (the distance recorded between a z-coordinate value of a carbon atom in the graphene sheet and the mean z-value of the D-glucose structure). The resultant system setup is depicted in Figure 7a.

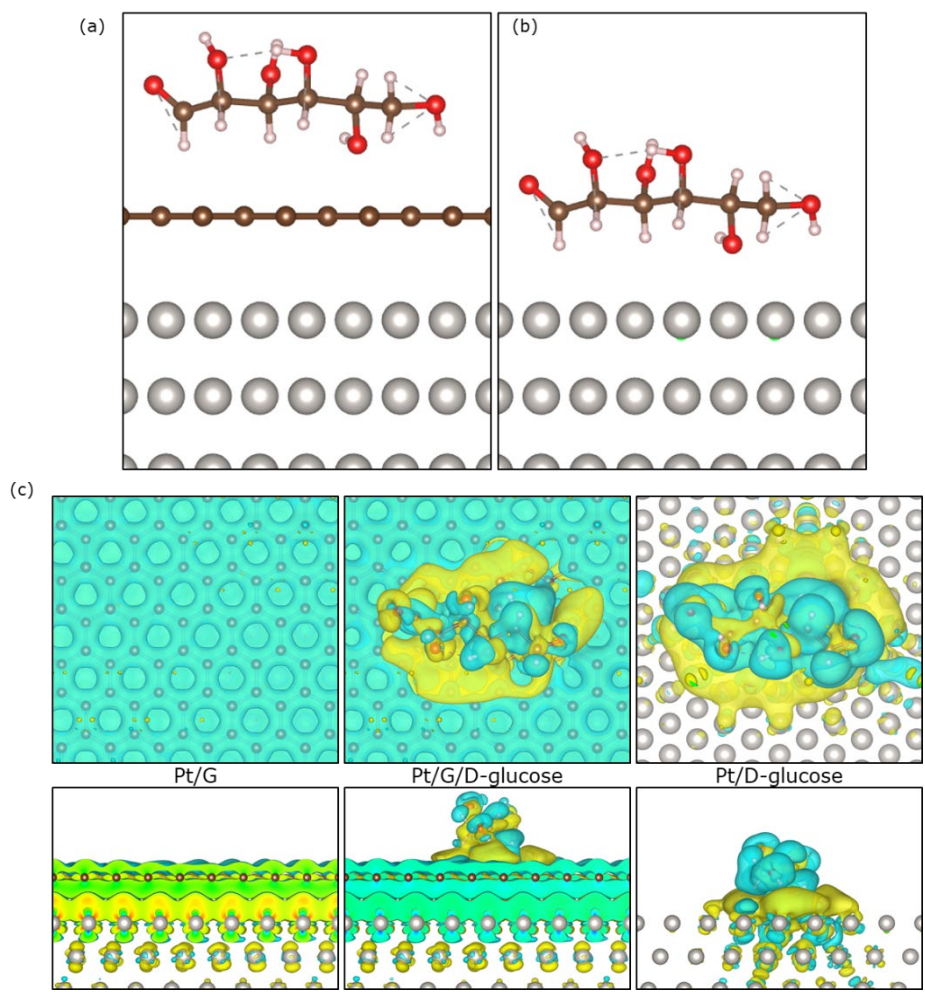

**Figure 7.** Systems setup. The left panel (**a**) depicts the adsorbed D-glucose on top of a relaxed graphene sheet on top of a Pt <111> substrate. The right panel (**b**) shows the relaxed D-glucose directly on top of the Pt111 substrate. The atoms are shown as spheres, where Pt atoms are rendered in grey, carbons in brown, oxygens in red and hydrogens in white; (**c**) Charge density difference contour plots (generated via VESTA package [VESTA]) for D-glucose on top of Pt <111> substrate and Pt <111>/G combined system. The top subfigures are top views for the contours, while the bottom subfigures show the side views. The atoms are shown as spheres.

The electronic charge density differences (CDDs) were calculated and extracted as 3D isosurface contour plots, as depicted in Figure 7 <CDDs>. The changes in the electronic charge densities were defined as indicated in Equation (1).

$$\Delta \rho = \rho_{Pt-G-glucose} - \rho_{Pt111} - \rho_G - \rho_{glucose} \tag{1}$$

From the CDDs, the charge depletion regions dominate across the graphene sheet while charge accumulation regions are concentrated on the top layer of the Pt substrate within the supercell sans the D-glucose adsorbate on top (as depicted in the left panel of Figure 7b. Adding the glucose on top, the charge depletion regions along the graphene sheet became thicker and associated with charge accumulation regions at the coordinates where glucose exists, while the charge accumulation regions alongside the top-most layer of the Pt substrate got thicker as well as depicted in the middle panel of CDDs Figure 7c. Meanwhile, in the relaxed D-glucose on top of the Pt substrate system, alternating charge accumulation and depletion regions moving from the Pt substrate towards the D-glucose were observed.

## 4. Conclusion

In this paper, we discussed a new graphene-heterostructured design and the applicability of using it for glucose level detection in diabetes monitoring. Fabrication and characterization techniques illustrated the working principle of the device by describing the changes in the Schottky barrier height. This was due to the presence of glucose molecules and the breakdown of gluconolactone due to the catalytic activity of Pt thin films. The presence of glucose molecules on the surface of the device resulted in a current that moves in the opposite direction to the applied bias (forward bias), with the difference between the two current values representing the concentration of glucose molecules. A selectivity test was obtained using different competitive solutions NaCl, KCl, sucrose, $Na_2SO_4$, and urea in two forms (1) separate and (2) mixed solutions to investigate the response of the device. It was noticed that there was a slight difference between the solutions, with a more noticeable difference when a droplet of glucose was added. An effect on Pt thickness was observed due to the presence of the same glucose concentration and due to the oxidation process. Density functional theory calculation results gave a qualitative explanation of the charges distributed across the graphene sheet within a system of the platinum substrate and D-glucose molecules on top. Further analysis will investigate the presence of other molecules in the presence of bias on the system. It will help in future efforts to further investigate the response of the device by quantum chemistry analysis.

**Author Contributions:** Conceptualization, M.S. and M.A.S.; methodology, M.S. and M.A.S.; software, M.A.S. and K.E.; validation, M.A.S., K.E., A.D. and M.S.; formal analysis, M.A.S. and M.S.; investigation, M.A.S., K.E. and M.S.; resources, M.S. and A.D.; data curation, M.A.S. and M.S.; writing—original draft preparation, M.A.S., M.S., K.E. and A.D.; writing—review and editing, M.A.S., M.S. and A.D.; visualization, M.A.S. and K.E.; supervision, A.D. (the computational section) and M.S. (the whole work); project administration, M.S.; funding acquisition, M.S. All authors have read and agreed to the published version of the manuscript.

**Funding:** This research was funded by an American University in Cairo (AUC) Faculty Support Grants# RSG-16 and RSG1-20. The computations were performed on resources provided by the Swedish National Infrastructure for Computing (SNIC) at the National Supercomputer Center (NSC), Linköping University, the PDC Centre for High Performance Computing (PDC-HPC), KTH, and the High Performance Computing Center North (HPC2N), Umeå University. For the computations, Anna Delin acknowledges financial support from Vetenskapsrådet, Swedish Energy Agency and the Knut and Alice Wallenberg foundation.

**Conflicts of Interest:** The authors declare no conflict of interest.

## References

1. Ferrari, M. (Ed.) *BioMEMS and Biomedical Nanotechnology: VI: Biomedical & Biological Nanotechnology. V2: Micro/Nano Technology for Genomics and Proteomics. V3: Therapeutic Micro/Nanotechnology. V4: Biomolecular Sensing, Processing and Analysis*; Springer-Verlag US, 2006.
2. Xie, X.; Zhao, W.; Lee, H.R.; Liu, C.; Ye, M.; Xie, W.; Cui, B.; Criddle, C.S.; Cui, Y. Enhancing the Nanomaterial Bio-Interface by Addition of Mesoscale Secondary Features: Crinkling of Carbon Nanotube Films To Create Subcellular Ridges *ACS Nano* **2014**, *8*, 11958–11965.
3. Rahmati, M.; Mozafari, M. Biological Response to Carbon-Family Nanomaterials: Interactions at the Nano-Bio Interface. *Front. Bioeng. Biotechnol.* **2019**, *7*, doi:10.3389/fbioe.2019.00004.
4. Marino, A.; Genchi, G.G.; Sinibaldi, E.; Ciofani, G. Piezoelectric Effects of Materials on Bio-Interfaces. *ACS Appl. Mater. Interfaces* **2017**, *9*, 17663–17680.
5. Global Blood Glucose Monitoring Market—Growth, Trends and Forecast (2019–2024). Available online: https://www.mordorintelligence.com/industry-reports/global-blood-glucose-monitoring-market-industry (accessed on 30 August 2019).
6. Lee, W.C.; Kim, K.B.; Gurudatt, N.G.; Hussain, K.K.; Choi, C.S.; Park, D.S.; Shim, Y.B. Comparison of enzymatic and non-enzymatic glucose sensors based on hierarchical Au-Ni alloy with conductive polymer. *Biosens. Bioelectron.* **2019**, *130*, 48–54.
7. Yuan, J.H.; Wang, K.; Xia, X.H. Highly Ordered Platinum-Nanotubule Arrays for Amperometric Glucose Sensing. *Adv. Funct. Mater.* **2005**, *15*, 803–809.

8. Thabit, H.; Lubina-Solomon, A.; Stadler, M.; Leelarathna, L.; Walkinshaw, E.; Pernet, A.; Allen, J.M.; Iqbal, A.; Choudhary, P.; Kumareswaran, K.; Nodale, M. Home use of closed-loop insulin delivery for overnight glucose control in adults with type 1 diabetes: A 4-week, multicentre, randomised crossover study. *Lancet Diabetes Endocrinol.* **2014**, *2*, 701–709.
9. Powell, D.R.; Smith, M.; Greer, J.; Harris, A.; Zhao, S.; DaCosta, C.; Mseeh, F.; Shadoan, M.K.; Sands, A.; Zambrowicz, B.; Ding, Z.M. LX4211 Increases Serum Glucagon-Like Peptide 1 and Peptide YY Levels by Reducing Sodium/Glucose Cotransporter 1 (SGLT1)–Mediated Absorption of Intestinal Glucose," *J. Pharmacol. Exp. Ther.* **2013**, *345*, 250–259.
10. Mei, L.; Zhang, P.; Chen, J.; Chen, D.; Quan, Y.; Gu, N.; Zhang, G.; Cui, R. Non-enzymatic sensing of glucose and hydrogen peroxide using a glassy carbon electrode modified with a nanocomposite consisting of nanoporous copper, carbon black and nafion. *Microchim. Acta* **2016**, *183*, 1359–1365.
11. Shenoy, R.; Tibbitt, M.W.; Anseth, K.S.; Bowman, C.N. Formation of Core–Shell Particles by Interfacial Radical Polymerization Initiated by a Glucose Oxidase-Mediated Redox System. *Chem. Mater.* **2013**, *25*, 761–767.
12. Zhao, S.; Mugabo, Y.; Iglesias, J.; Xie, L.; Delghingaro-Augusto, V.; Lussier, R.; Peyot, M.L.; Joly, E.; Taïb, B.; Davis, M.A.; Brown, J.M. $\alpha/\beta$-Hydrolase domain-6-accessible monoacylglycerol controls glucose-stimulated insulin secretion. *Cell Metab.* **2014**, *19*, 993–1007.
13. Toghill, K.E.; Compton, R.G. Electrochemical non-enzymatic glucose sensors:a perspective and an evaluation. *Int. J. Electrochem. Sci.* **2010**, *5*, 1246–1301.
14. Kwon, S.Y.; Kwen, H.D.; Choi, S.H. Fabrication of Nonenzymatic Glucose Sensors Based on Multiwalled Carbon Nanotubes with Bimetallic Pt-M (M = Ru and Sn) Catalysts by Radiolytic Deposition. *J. Sens.* **2012**, *2012*, 784167. Available online: https://www.hindawi.com/journals/js/2012/784167/ (accessed on 7 November 2019).
15. Tortorich, R.P.; Shamkhalichenar, H.; Choi, J.W. Inkjet-Printed and Paper-Based Electrochemical Sensors. *Appl. Sci.* **2018**, *8*, 288.
16. Scandurra, A.; Ruffino, F.; Sanzaro, S.; Grimaldi, M.G. Laser and thermal dewetting of gold layer onto graphene paper for non-enzymatic electrochemical detection of glucose and fructose. *Sens. Actuators B Chem.* **2019**, *301*, 127113.
17. Juřík, T.; Podešva, P.; Farka, Z.; Kováŕ, D.; Skládal, P.; Foret, F. Nanostructured gold deposited in gelatin template applied for electrochemical assay of glucose in serum. *Electrochimica Acta* **2016**, *188*, 277–285.
18. Bai, Z.; Li, G.; Liang, J.; Su, J.; Zhang, Y.; Chen, H.; Huang, Y.; Sui, W.; Zhao, Y. Non-enzymatic electrochemical biosensor based on Pt NPs/RGO-CS-Fc nano-hybrids for the detection of hydrogen peroxide in living cells. *Biosens. Bioelectron.* **2016**, *82*, 185–194.
19. Shi, J.J.; Hu, W.; Zhao, D.; He, T.T.; Zhu, J.J. Sonoelectrochemical synthesized RGO–PbTe composite for novel electrochemical biosensor. *Sens. Actuators B Chem.* **2012**, *173*, 239–243.
20. Wu, G.H.; Song, X.H.; Wu, Y.F.; Chen, X.M.; Luo, F.; Chen, X. Non-enzymatic electrochemical glucose sensor based on platinum nanoflowers supported on graphene oxide. *Talanta* **2013**, *105*, 379–385.
21. Rengaraj, A.; Haldorai, Y.; Kwak, C.H.; Ahn, S.; Jeon, K.J.; Park, S.H.; Han, Y.K.; Huh, Y.S. Electrodeposition of flower-like nickel oxide on CVD-grown graphene to develop an electrochemical non-enzymatic biosensor. *J. Mater. Chem. B* **2015**, *3*, 6301–6309.
22. Yu, X.; Zhang, Y.; Guo, L.; Wang, L. Macroporous carbon decorated with dendritic platinum nanoparticles: One-step synthesis and electrocatalytic properties. *Nanoscale* **2014**, *6*, 4806–4811.
23. Chun, S.; Kim, Y.; Oh, H.-S.; Bae, G.; Park, W. A highly sensitive pressure sensor using a double-layered graphene structure for tactile sensing. *Nanoscale* **2015**, *7*, 11652–11659.
24. Shafiei, M.; Spizzirri, P.G.; Arsat, R.; Yu, J.; du Plessis, J.; Dubin, S.; Kaner, R.B.; Kalantar-Zadeh, K.; Wlodarski, W. Platinum/Graphene Nanosheet/SiC Contacts and Their Application for Hydrogen Gas Sensing. *J. Phys. Chem. C* **2010**, *114*, 13796–13801.
25. Chaves, F.A.; Jiménez, D.; Cummings, A.W.; Roche, S. Model of the Electrostatics and Tunneling Current of Metal-Graphene Junctions and Metal-Insulator-Graphene Heterostructures. *ArXiv* **2013**, ArXiv13090390.
26. Abdelnasser, S.; Sakr, M.A.; Serry, M. Nanostructured graphene-platinum-PEDOT electrode materials for enhanced Schottky performance and power conversion applications. *Microelectron. Eng.* **2019**, *216*, 111045.
27. Ye, J.S.; Hong, B.D.; Wu, Y.S.; Chen, H.R.; Lee, C.L. Heterostructured palladium-platinum core-shell nanocubes for use in a nonenzymatic amperometric glucose sensor. *Microchim. Acta* **2016**, *183*, 3311–3320.
28. Zeghbroeck, B.V. *Principles of Semiconductor Devices and Heterojunctions*; Prentice Hall PTR: Upper Saddle River, NJ, USA, 2007.

29. Sakr, M.A.; Serry, M. Non-enzymatic graphene-based biosensors for continous glucose monitoring. In Proceedings of the 2015 IEEE SENSORS, Busan, South Korea, 1–4 November 2015; pp. 1–4.
30. Badhulika, S.; Paul, R.K.; Terse, T.; Mulchandani, A. Nonenzymatic Glucose Sensor Based on Platinum Nanoflowers Decorated Multiwalled Carbon Nanotubes-Graphene Hybrid Electrode. *Electroanalysis* **2014**, *26*, 103–108.
31. Seah, C.M.; Chai, S.P.; Mohamed, A.R. Mechanisms of graphene growth by chemical vapour deposition on transition metals. *Carbon* **2014**, *70*, 1–21.
32. Hsieh, Y.-P.; Hofmann, M.; Kong, J. Promoter-assisted chemical vapor deposition of graphene. *Carbon* **2014**, *67*, 417–423.
33. Serry, M.; Sakr, M.A. Study of Flexoelectricity in Graphene Composite Structures. *MRS Adv.* **2016**, *1*, 2723–2729.
34. Serry, M.; Sakr, M.A. Modeling and experimental characterization of flexible graphene composite strain sensors. In Proceedings of the 2016 IEEE SENSORS, Orlando, FL, USA, 30 October–3 November 2016; pp. 1–3.
35. Tian, K.; Prestgard, M.; Tiwari, A. A review of recent advances in nonenzymatic glucose sensors. *Mater. Sci. Eng. C* **2014**, *41*, 100–118.
36. Giannozzi, P.; Baroni, S.; Bonini, N.; Calandra, M.; Car, R.; Cavazzoni, C.; Ceresoli, D.; Chiarotti, G.L.; Cococcioni, M.; Dabo, I.; Dal Corso, A. QUANTUM ESPRESSO: A modular and open-source software project for quantum simulations of materials. *J. Phys. Condens. Matter Inst. Phys. J.* **2009**, *21*, 395502.
37. Björkman, T. CIF2Cell: Generating geometries for electronic structure programs," *Comput. Phys. Commun.* **2011**, *182*, 1183–1186.
38. Hamann, D.R. Optimized norm-conserving Vanderbilt pseudopotentials. *Phys. Rev. B* **2013**, *88*, 085117.
39. Gilbert, A. IQmol is an Open Source Molecular Editor and Visualization Package; 2017. Available online: http://www.iqmol.org/index.html (accessed on 24 December 2019)
40. Al-Gaashani, R.; Najjar, A.; Zakaria, Y.; Mansour, S.; Atieh, M.A. XPS and structural studies of high quality graphene oxide and reduced graphene oxide prepared by different chemical oxidation methods. *Ceram. Int.* **2019**, *45*, 14439–14448.
41. Al Balushi, Z.Y.; Wang, K.; Ghosh, R.K.; Vilá, R.A.; Eichfeld, S.M.; Caldwell, J.D.; Qin, X.; Lin, Y.C.; DeSario, P.A.; Stone, G.; Subramanian, S. Two-dimensional gallium nitride realized via graphene encapsulation. *Nat. Mater.* **2016**, *15*, 1166–1171.
42. Giovannetti, G.; Khomyakov, P.A.; Brocks, G.; Karpan, V.M.; van den Brink, J.; Kelly, P.J. Doping Graphene with Metal Contacts. *Phys. Rev. Lett.* **2008**, *101*, 026803.
43. Chu, B.H.; Lo, C.F.; Nicolosi, J.; Chang, C.Y.; Chen, V.; Strupinski, W.; Pearton, S.J.; Ren, F. Hydrogen detection using platinum coated graphene grown on SiC. *Sens. Actuators B Chem.* **2011**, *157*, 500–503.
44. Rochefort, A.; Yang, D.Q.; Sacher, E. Stabilization of platinum nanoparticles on graphene by non-invasive functionalization. *Carbon* **2009**, *47*, 2233–2238.
45. Grimme, S. Semiempirical GGA-type density functional constructed with a long-range dispersion correction. *J. Comput. Chem.* **2006**, *27*, 1787–1799.
46. Barone, V.; Casarin, M.; Forrer, D.; Pavone, M.; Sambi, M.; Vittadini, A. Role and effective treatment of dispersive forces in materials: Polyethylene and graphite crystals as test cases. *J. Comput. Chem.* **2009**, *30*, 934–939.
47. May, P.; Lazzeri, M.; Venezuela, P.; Herziger, F.; Callsen, G.; Reparaz, J.S.; Hoffmann, A.; Mauri, F.; Maultzsch, J. Signature of the two-dimensional phonon dispersion in graphene probed by double-resonant Raman scattering. *Phys. Rev. B* **2013**, *87*, 075402.
48. Ferrari, A.C. Raman spectroscopy of graphene and graphite: Disorder, electron–phonon coupling, doping and nonadiabatic effects. *Solid State Commun.* **2007**, *143*, 47–57.
49. No, Y.S.; Choi, H.K.; Kim, J.S.; Kim, H.; Yu, Y.J.; Choi, C.G.; Choi, J.S. Layer number identification of CVD-grown multilayer graphene using Si peak analysis. *Sci. Rep.* **2018**, *8*, 571.
50. Gao, H.; Xiao, F.; Ching, C.B.; Duan, H. One-Step Electrochemical Synthesis of PtNi Nanoparticle-Graphene Nanocomposites for Nonenzymatic Amperometric Glucose Detection. *ACS Appl. Mater. Interfaces* **2011**, *3*, 3049–3057.
51. Kuila, T.; Bose, S.; Khanra, P.; Mishra, A.K.; Kim, N.H.; Lee, J.H. Recent advances in graphene-based biosensors. *Biosens. Bioelectron.* **2011**, *26*, 4637–4648.
52. Dan, Y.; Lu, Y.; Kybert, N.J.; Luo, Z.; Johnson, A.T.C. Intrinsic Response of Graphene Vapor Sensors. *Nano Lett.* **2009**, *9*, 1472–1475.

53. Chekin, F.; Singh, S.K.; Vasilescu, A.; Dhavale, V.M.; Kurungot, S.; Boukherroub, R.; Szunerits, S. Reduced Graphene Oxide Modified Electrodes for Sensitive Sensing of Gliadin in Food Samples. *ACS Sens.* **2016**, *1*, 1462–14706.
54. Yang, J.; Kwak, T.J.; Zhang, X.; McClain, R.; Chang, W.J.; Gunasekaran, S. Digital pH Test Strips for In-Field pH Monitoring Using Iridium Oxide-Reduced Graphene Oxide Hybrid Thin Films. *ACS Sens.* **2016**, *1*, 1235–1243.
55. Anota, E.C.; Soto, A.T.; Cocoletzi, G.H. Studies of graphene–chitosan interactions and analysis of the bioadsorption of glucose and cholesterol. *Appl. Nanosci.* **2014**, *4*, 911–918.
56. Wehling, T.O.; Lichtenstein, A.I.; Katsnelson, M.I. First-principles studies of water adsorption on graphene: The role of the substrate. *Appl. Phys. Lett.* **2008**, *93*, 202110.
57. Subrahmanyam, K.S.; Manna, A.K.; Pati, S.K.; Rao, C.N.R. A study of graphene decorated with metal nanoparticles. *Chem. Phys. Lett.* **2010**, *497*, 70–75.
58. Smith, A.D.; Elgammal, K.; Fan, X.; Lemme, M.; Delin, A.; Niklaus, F.; Östling, M. Toward effective passivation of graphene to humidity sensing effects. In Proceedings of the 2016 46th European Solid-State Device Research Conference (ESSDERC), Lausanne, Switzerland, 12–15 September 2016; pp. 299–302.
59. Smith, A.D.; Elgammal, K.; Niklaus, F.; Delin, A.; Fischer, A.C.; Vaziri, S.; Forsberg, F.; Råsander, M.; Hugosson, H.; Bergqvist, L.; Schröder, S. Resistive graphene humidity sensors with rapid and direct electrical readout. *Nanoscale* **2015**, *7*, 19099–19109.
60. Elgammal, K.; Hugosson, H.W.; Smith, A.D.; Råsander, M.; Bergqvist, L.; Delin, A. Density functional calculations of graphene-based humidity and carbon dioxide sensors: Effect of silica and sapphire substrates. *Surf. Sci.* **2017**, *663*, 23–30.
61. Smith, A.D.; Elgammal, K.; Fan, X.; Lemme, M.C.; Delin, A.; Råsander, M.; Bergqvist, L.; Schröder, S.; Fischer, A.C.; Niklaus, F.; et al. Graphene-based $CO_2$ sensing and its cross-sensitivity with humidity. *RSC Adv.* **2017**, *7*, 22329–22339.